\documentstyle[prl,aps,epsfig,multicol]{revtex}

\def\be{\begin{equation}}

\def\j-{\J_-}

\def\ee{\end{equation}}

\def\be{\begin{equation}}
\def\ee{\end{equation}}

\def\bea{\begin{eqnarray}}
\def\eea{\end{eqnarray}}
\def\bearr{\begin{eqnarray}}
\def\bearrs{\begin{eqnarray*}}
\def\eearr{\end{eqnarray}}
\def\eearrs{\end{eqnarray*}}
\def\barr{\begin{array}}
\def\earr{\end{array}}

\def\non\non{\nonumber}
\def\nn8{\nonumber\\[15pt]}

\def\gev{{\rm \,Ge\kern-0.125em V}}
\def\mpc{{\rm \,Mpc}}

\def\calT{{\cal T}}
\def\calR{{\cal R}}

\def\ln{{\rm{ln}}}
\newcommand\lsim{\mathrel{\rlap{\lower4pt\hbox{\hskip1pt$\sim$}}
    \raise1pt\hbox{$<$}}}
\newcommand\gsim{\mathrel{\rlap{\lower4pt\hbox{\hskip1pt$\sim$}}
    \raise1pt\hbox{$>$}}}
\begin{document}
\title{Temperature of the inflaton and duration of inflation from WMAP data  }
\author{Kaushik Bhattacharya,
 Subhendra Mohanty and Raghavan Rangarajan }
\address{ Physical Research Laboratory,
Navrangpura, Ahmedabad - 380 009, India}
\date{\today}

\maketitle

\begin{abstract}
If the initial state of the inflaton field is taken to have a thermal
distribution instead of the conventional zero particle vacuum state
then the curvature power spectrum gets modified by a temperature
dependent factor such that the fluctuation spectrum of the microwave
background radiation is enhanced at larger angles. We compare this
modified cosmic microwave background spectrum with Wilkinson microwave
anisotropy probe data to obtain an upper bound on the temperature of
the inflaton at the time our current horizon crossed the horizon
during inflation. We further conclude that there must be additional
e-foldings of inflation beyond what is needed to solve the horizon
problem.
\end{abstract}
%
\begin{multicols}{2}
The inflationary paradigm \cite{inflation}
provides a ready solution for the horizon and flatness problems while
simultaneously providing a source for the initial density
perturbations required as a seed for the large scale structure in the
universe today.  During inflation the energy density of the universe
is dominated by the potential energy of a scalar field leading to a
brief period of accelerated expansion.

The amplitude for quantum fluctuations in the inflaton field $\phi$
generated during inflation effectively freezes out for a particular
comoving scale once that mode crosses the horizon.  These fluctuations
also translate into perturbations in energy density and curvature in
the universe.  In the inflationary paradigm these perturbations are
the source of the observed anisotropy in the microwave background and
the seed for large scale structure.  Conventionally the fluctuation
power spectrum for the inflaton is calculated using the zero particle
vacuum. However if inflation was preceded by a radiation era then the
inflaton was in thermal equilibrium at some point in the early
universe, via at least gravitational couplings at the Planck scale.
In chaotic \cite{chaotic} and natural inflation models with a symmetry
breaking scale $f\sim M_{Pl}$ \cite{natural}, non-zero
momentum modes can be in thermal equilibrium due to gravitational
interactions with the existing radiation.  The inflaton (or any other
field which contributes to the density perturbation during inflation)
can then have a thermal distribution corresponding to the temperature
${\cal T}_{i}$ at the beginning of inflation.  Though the inflaton
field may be decoupled from the rest of radiation before inflation it
retains its thermal distribution in an adiabatically expanding
universe. (This is different from the warm inflation scenarios
\cite{warm} where the inflaton decays to radiation during inflation.)
In reference \cite{venenziano} it was argued that the power spectrum
of gravitational waves generated during inflation has an extra
temperature dependent factor due to the thermal gravitons which
decouple at the Planck era and retain their thermal distribution.

In this Letter we include the initial thermal nature of the inflaton
$\phi$ in the generation of the inflaton fluctuation power spectrum.
The power spectrum of the perturbations of the inflaton, and hence
of the curvature, has an extra temperature dependent factor $\coth
[k/(2 a_i{\cal T}_{i})]$ where $k$ is the wavenumber of the modes in
comoving coordinates and $a_i$ is the scale factor at $t_i$ when
inflation commences.  The physical wavenumber is $k/a$ and physical
temperature ${\cal T}=T/a$, where $T$ is the comoving temperature
during inflation.  Using this thermal power spectrum we calculate
the CMB anisotropy using CMBFAST \cite{cmbfast}. We compare the
result with data from WMAP \cite{wmapdata} and find that the
comoving temperature of the inflaton is constrained to be $ T < 1.0
\times 10^{-3}\mpc^{-1}$.  Since $T=\calT_0a_0$, where $\calT_0$ and
$a_0$ are the temperature and scale factor when our current horizon
scale crossed the horizon during inflation, this constraint can be
rewritten as $\calT_0<4.2 H$, where $H$ is the Hubble parameter
during inflation (all bounds stated in this paper are at $1 \sigma$
or $68 \% C.L.$). This result is valid independent of the scale of
inflation. If inflation takes place in the GUT era ($M_{GUT} \sim
10^{15} \gev$) then ${\cal T}_0 < 1.0 \times 10^{12} \gev$ and for
the inflation at the electroweak era ($M_{EW} \sim 10^{3} \gev$)
\cite{Knox:1992iy} we have the bound ${\cal T}_0 < 1.0 \times
10^{-12} \gev$.
 Since $\calT_0$ is much less than the energy scale
of inflation, $M_{inf}$, the duration of inflation must be longer than what is
needed to solve the horizon problem in models where the temperature
at the beginning of inflation $\calT_i \sim M_{inf}/7$, the energy
scale of inflation.  If $\Delta N$ is the number of e-foldings from
the beginning of inflation to the time our current horizon scale
crossed the de Sitter horizon, then the bound on $\calT_0$ from WMAP
data implies $\Delta N =\ln(\calT_i/\calT_0)>\ln(0.03 \,M_{inf}/H)$.
For GUT scale ($M_{inf}=10^{15} \gev$) inflation $\Delta N_{GUT} >
7$ and for inflation at the the electroweak scale $\Delta N_{EW} >
32$.

We now calculate the 
curvature power spectrum due to inflaton fluctuations.
Presuming a quasi de Sitter
universe during inflation, conformal time $\tau$ ($d\tau \equiv dt/a$)
and the scale factor during inflation $a(\tau)$ are related by
$a(\tau) = -1/H\tau(1- \epsilon)$ where $\epsilon
=-\frac{\dot{H}}{H^2}=\frac{4\pi}{M_{Pl}^2}\frac{\dot\phi^2}{H^2}
=\frac{M_{Pl}^2}{16\pi}\left(\frac{V'}{V} \right)^2$, $V$ is the
inflaton potential, $V'$ stands for the derivative of the potential
with respect to $\phi$ and $\dot{H}$ implies a derivative of $H$ with
respect to cosmic time. Henceforth a dot will always signify a
derivative with respect to cosmic time whereas a prime may mean a
derivative with respect to the conformal time when it appears with
quantities which are explicit functions of time or a derivative with
respect to the inflaton field when it appears with the
potential. $M_{Pl}$ is the Plank mass. We define additional slow roll
parameters as $\eta=\frac{M^2_{Pl}}{8\pi}\left(\frac{V''}{V}\right)$
and $\delta=\ddot\phi/(H\dot\phi)=\epsilon-\eta$.  We further define
$z=\frac{a \dot{\phi}}{H}$ and $u=a(\tau)\delta\phi(\bf{x},\tau)$,
where $\delta\phi(\bf{x},\tau)$ is the inflaton field perturbations on
spatially flat hypersurfaces.  Then the action for $u$ is given as
\cite{mukha1,mukha2,Stewart:1993bc}:
\begin{eqnarray}
S &=&\frac12\int d\tau \,d^3x\left[(u')^2 - ({\bf{\nabla}}\,u)^2
+\frac{z''}{z}u^2\right]\,,
\end{eqnarray}
The gauge invariant comoving curvature perturbation is defined as
${\cal R}=\psi - H\frac{\delta \phi}{\dot{\phi}}$ where $\psi$ is
the usual scalar metric perturbation. Consequently on spatially flat
hypersurfaces $u=-z{\cal R}$. Expressing $u({\bf x},\tau)$ as a
quantum field we can write:
\begin{eqnarray}
\hat u({\bf x},\tau) &=& \int \frac{d^3 k} {(2\pi)^{3/2}}
\Bigl(a_{\bf k} ~f_k(\tau) +a^{\dagger}_{-{\bf k}}~f_{k}^*(\tau)
\Bigr)\,e^{i{\bf  k} \cdot {\bf x}}\,, \label{fe}
\end{eqnarray}
where ${\bf k}$ is the comoving wavenumber.

The gauge invariant comoving curvature perturbation can then be expressed as:
\bea {\cal R}&=&\frac{-1}{z}\int \frac{d^3 k} {(2\pi)^{3/2}}
\Bigl(a_{\bf k} ~f_k(\tau) +a^{\dagger}_{-{\bf k}}~f_{k}^*(\tau)
\Bigr)\,e^{i{\bf  k} \cdot {\bf x}}\,, \nonumber\\
&\equiv& \int \frac{d^3 k}{(2\pi)^{3/2}}{\cal R}_{\bf
k}(\tau)\,e^{i{\bf k} \cdot {\bf x}} \,.
\label{ftR}
\eea
The power spectrum of the comoving curvature
perturbations can be defined by the relation
\be
\langle {\cal
R}_{\bf k} {\cal R}^*_{\bf k} \rangle \equiv \frac{2 \pi^2}{k^3}
P_{\cal R} \, \delta^3(\bf k -\bf k^\prime)
\label{power}
\ee
The usual quantization condition between the fields and their
canonical momenta yields $[a_{\bf k} , a^\dagger_{\bf k^\prime}]=
\delta^3({\bf k-k^\prime})$ and the vacuum satisfies $a_{\bf
k}|0\rangle=0$.  If the inflaton field had zero occupation prior to
inflation then $ \langle a_{\bf k}^\dagger a_{\bf k}\rangle=0$ and we
would obtain a correlation function $\sim |f_k(\tau)|^2$. However if
the inflaton field was in thermal equilibrium at some earlier epoch
\cite{guthpi} it will retain its thermal distribution even after
decoupling from the other radiation fields and its occupation number
will be given by:
\be \langle a_{\bf k}^\dagger a_{{\bf k}'}\rangle =
\frac{1}{e^{E_{k} /\calT_f} -1} \delta^3({\bf k}-{\bf k}')\, ,
\label{thermal} \ee
where $E_{k}$ is the energy corresponding to the $k$ mode at the
inflaton decoupling temperature $\calT_f$. For effectively free
modes $E_k\approx k/a_f$, and so $E_{k}/\calT_f=k/(a_f\calT_f)=
k/(a_i\calT_i)=k/T$.

Using Eq.~(\ref{ftR}) and Eq.~(\ref{thermal}) it can be seen that
\bea
\langle {\cal
R}_{\bf k} {\cal R}^*_{\bf k} \rangle &=& \frac{1}{|z|^2}
\,\left(1 + \frac{2}{e^{\frac{k}{T}}-1}\right)|f_k(\tau)|^2
\, \delta^3(\bf k -\bf k^\prime)
\nonumber\\
&=& \frac{1} {|z|^2}|f_k(\tau)|^2\,\coth\left[\frac{k}{2
T}\right]\, \delta^3(\bf k -\bf k^\prime) \label{R2}
\eea
From the defining relation Eq.~(\ref{power}) for the curvature power
spectrum and Eq.~(\ref{R2}) we find that the power spectrum for the
curvature perturbations can be expressed in terms of the mode functions
$f_k(\tau)$ as
\be
P_{\cal
R}(k)=\frac{k^3}{2\pi^2}\frac{|f_{k}|^2}{|z|^2}
\,\coth\left[\frac{k}{2 T}\right]
\label{power2}
\ee
For constant $\epsilon$ and $\delta$ the mode functions
$f_k(\tau)$ obey the minimally coupled Klein-Gordon equation
\cite{Stewart:1993bc,riotto},
\be
f_k^{\prime \prime} + \left[k^2 - \frac{1}{\tau^2}\left(\nu^2 -
\frac14\right) \right]f_k=0\,,
\label{f2}
\ee
where $k=|{\bf k}|$ and, for small $\epsilon$ and $\delta$,
$\nu=\frac32 + 2\epsilon +\delta$. Eq.~(\ref{f2}) has the
general solution given by,
\be
f_k( \tau)=\sqrt{-\tau}\left[c_1(k)\,H^{(1)}_\nu(-k\tau)+
c_2(k)\,H^{(2)}_\nu(-k\tau)\right]\,.
\label{sol1}
\ee
When the modes are well within the horizon they can be
approximated by flat spacetime solutions ${f_k}^0(\tau) =
\frac{1}{\sqrt{2 k}}e^{-i k \tau}\,, ~~~~~(k \gg  a H)\,. $
Matching the general solution in Eq.~(\ref{sol1}) with the
solution in the high frequency (``flat spacetime") limit gives
the value of the constants of integration $
c_1(k)=\frac{\sqrt{\pi}}{2}e^{i(\nu +
\frac12)\frac{\pi}{2}}\,~~~{\rm and}~~~~c_2(k) = 0\,.$
Eq.~(\ref{sol1}) then implies that for $-k\tau \gg 1$ or $k \ll aH$,
\be f_k(\tau)=e^{i(\nu - \frac12)\frac{\pi}{2}}2^{\nu-\frac32}
\frac{\Gamma(\nu)}{\Gamma(\frac32)}\frac{1}{\sqrt{2k}}(-k\tau)^{\frac12-\nu}\,.
\label{superhorizon}
\ee
Substituting the solution in Eq.~(\ref{superhorizon}) for the the
super-horizon modes ($k\ll aH$) in the expression Eq.~(\ref{power2})
for the curvature power spectrum we obtain
\be P_{\cal R}(k) =\frac
{H^4}{4\pi^2\dot\phi^2}\,\left(\frac{k}{aH}\right)^{n_s-1}
\,\coth\left[\frac{k}{2 T}\right]
\label{PR2}\\
\ee
with $n_s-1=3-2\nu=-4\epsilon-2\delta$. Note that since
$f_{k}(\tau)\sim z(\tau)$ for superhorizon modes, $P_{\cal R}(k)$ is
time independent.

We can now rewrite the power spectrum as,
\be P_{\calR}(k) = A(k_0)~\left(\frac{k}{k_0}\right)^{n_s-1}~
\coth\left[\frac{k}{2 T}\right]\, \label{pR} \ee
where $k_0$ is referred to as the pivot point and $A(k_0)$ is the
normalisation constant.
\be
A(k_0)=\frac
{H^4}{4\pi^2\dot\phi^2}\big|_{aH=k_0} =\frac{H_{k_0}^2}{\pi
M_{Pl}^2\epsilon}\, ,
\ee
where $H_{k_0}$ is the Hubble parameter
evaluated when $aH=k_0$ during inflation.

During inflation a mode $k$ leaves the horizon when $k/a =H$.
Therefore for any $k$ the initial thermal effects modify the
scale-free power spectrum by $\coth(a_k H_k/2 T)=\coth(H_k/2{\cal
T}_k)$, where $a_k$ and ${\cal T}_k$ are the
scale factor and
temperature when the mode $k$ leaves the horizon.

We now use the curvature power spectrum as given in Eq.~(\ref{pR}) to
obtain the CMB fluctuations.  We make use of the publicly available
code CMBFAST \cite{cmbfast} and modify the power spectrum formula,
according to our need, in subroutine powersflat in the program
cmbflat.F.  We set the parameters $\Omega_b=0.046, \Omega_m=0.27,
h=71.0$ as determined in \cite{spergel} for a flat universe with no
running of the scalar index $n_s$.  We set $k_0=0.05\mpc^{-1}$.

Fig. 1 shows two different possible CMB TT-correlation curves for two
comoving inflaton temperatures, with $n_s=0.99$ and $A(k_0)=2.01
\times 10^{-9}$.  It shows clearly that as the comoving temperature of
the inflatons decreases we get a better fit of our TT-correlations
with the WMAP data. The curves show that the change of temperature $T$
of the inflatons essentially affects the lower multipoles or higher
length scales of the TT-correlations while the higher multipoles
remains unaffected. This is expected because $\coth(x)\sim 1$ for
$x\gg1$.

We now do a 3-parameter analysis by varying $A(k_0)$ and $T$ for
multiple values of $n_s$ from $n_s=0.90$ to $n_s=1.1$, and each time
we compare the resultant TT-anisotropy with the WMAP data
\cite{wmapdata} and calculated the $\chi^2$ of the resultant computed
distribution with respect to the WMAP data. From this three parameter fit we find  that $\chi^2$
minimum occurs at the values $A(k_0)=2.0 \times 10^{-9}$, $n_s=0.99$
and at the comoving temperature of the inflaton field $T=0$.
In Fig. 2 we show the $1\sigma$ allowed regions of $T$ and $A(k_0)$
at different values of $n_s$. The regions inside the curves are
allowed by WMAP data, with the best fit value being $A(k_0)= 2.0\times
10^{-9}$, $n_s=0.99$, and comoving temperature $T=0$. The
maximum allowed values for $T$ increases when we increase $n_s$ from
$n_s=0.90$ to $n_s=1.06$. For values of $n_s$ greater than $1.06$
maximum allowed value of $T$ decreases monotonically. So from this
three parameter fits we see that the upper bound on the comoving
temperature of the inflaton field is $T< 0.001 \mpc^{-1}$.

The power spectrum of tensor perturbations also has an identical
$\coth(k/2T)$ factor due to the thermal spectrum of the gravitons in
the initial state \cite{venenziano}. The tensor-to-scalar amplitude
ratio $r=16 \epsilon <1.28\, (95 \% C.L.)$ \cite{peiris}. In our
parameter fits we do not violate this upper bound on $ \epsilon$ which
is independent of temperature.

The temperature of the inflaton $(\calT_0)$ when our present
horizon was leaving the de Sitter horizon $(H^{-1})$ can be
calculated, $T=\calT_0 a_0< 4.2 R_h^{-1}$ (where $a_0$ is the scale
factor at the time of this horizon crossing). We then obtain
\be {\cal T}_0 < 4.2 \left(\frac {a_{now}}{a_0}\right)R_h^{-1}=4.2 H
\, ,\label{T0}
\ee
where $a_{now}$ is the current scale factor and $R_h/a_{now}= H^{-1}/a_0$.
$H =(8 \pi/3)^{1/2} M_{inf}^2/M_{Pl}$, ignoring the variation in $H$ during
inflation. The constraint Eq.~(\ref{T0}) is
valid whatever be the scale of inflation. If inflation takes place
in the GUT era ($M_{GUT} \sim 10^{15} \gev$) then ${\cal T}_0 < 1.0
\times 10^{12} \gev$, and for inflation at the electroweak scale
($M_{EW} \sim 10^{3} \gev$) we have the bound ${\cal T}_0 < 1.0
\times 10^{-12} \gev$.
In general, for $M_{inf} <1.4\times 10^{17}\gev$,
$\calT_0<\calT_i$.
\begin{figure}[t]
\begin{center}
\epsfig{file=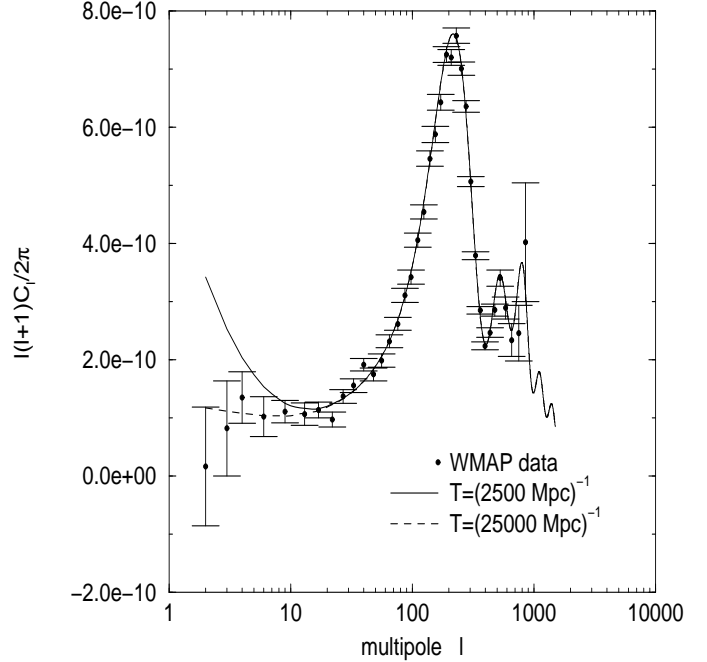,angle=0,width=9cm,height=9cm}
\end{center}
\caption[]
{\small\sf The dotted points with the error bars corresponds to the WMAP
binned data for the CMB TT-anisotropies.
The solid and dashed curves
are generated from CMBFAST with the inflaton comoving temperature
$T=(2500 {\rm Mpc})^{-1}$ and $T=(25000 {\rm Mpc})^{-1}$ respectively.
The error bars on the WMAP data consists of measurement errors and errors
attributed to cosmic variance.}
\label{f:ps}
\end{figure}
From Fig. 2 we see that $T$ has an upper bound at $1\sigma$ given by
\be T < 1.0 \times 10^{-3} {\mpc}^{-1}= 4.2 R_h^{-1}\,\,( 68.3 \%
C.L.)
\ee
where we have expressed the bound on the comoving temperature in terms
of the present horizon scale $R_h \simeq 4200 \mpc$.

The minimum number of e-foldings of inflation required to solve
the horizon problem
is obtained by taking our current
horizon scale to be the first to cross the horizon during
inflation.
If there is a thermal
background of the inflaton   and inflation commences at ${\cal
T}_i \sim M_{inf}/7$ our results above imply that our current
horizon must have crossed the de Sitter horizon some time after
the beginning of inflation.  Thus the duration of inflation must
be longer than what is just needed to solve the horizon problem
by $\Delta N =\ln({\cal T}_i/{\cal T}_0)$.
\begin{figure}
\begin{center}
\epsfig{file=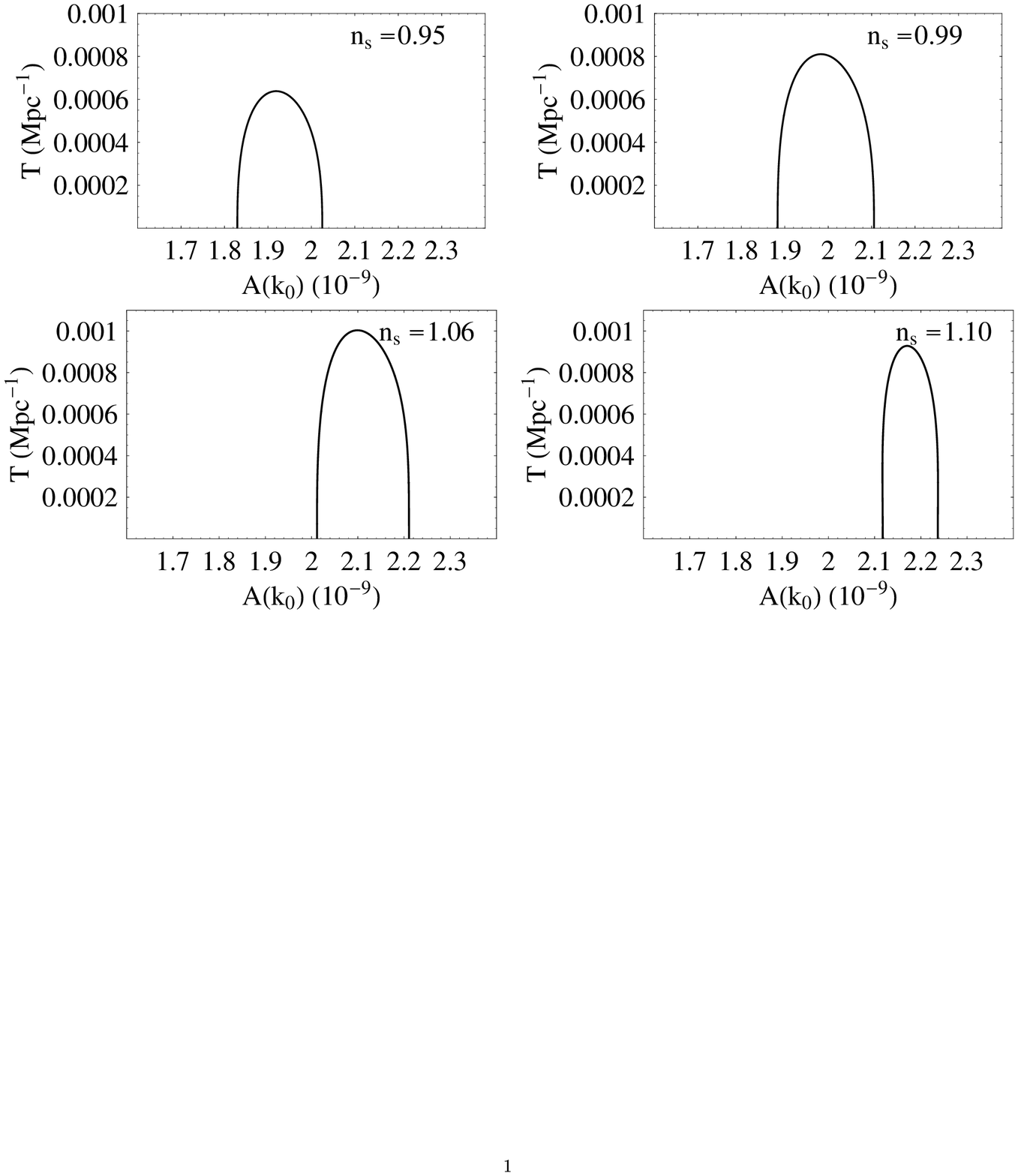,angle=0,width= 8.5cm,height=10cm}
\end{center}
\caption[] {\small\sf Values of $T$ and $A(k_0)$ allowed at
$1\sigma$ by WMAP data are the regions inside the curves. The plots
at different values of $n_s$ show that the maximum allowed value of
$T$ is $0.001 \mpc^{-1}$. } \label{f:chis}
\end{figure}
The upper bound on
${\cal T}_0$ from WMAP then implies
\be \Delta N >
\ln\left[0.03 \frac{M_{inf}}{H}\right]=\ln\left[ 0.01
\frac{M_{Pl}}{ M_{inf}}\right] \, . \label{dN} \ee
%
Perturbations generated during these additional e-foldings
are yet to enter
our present horizon.
 For GUT scale inflation WMAP data implies $\Delta
N_{GUT}>7$ and for inflation at the the electroweak scale it
implies $\Delta N_{EW}>32$. 
For Planck scale inflation, such as chaotic and natural inflation, no additional
e-foldings of inflation are required.

There exists a Lyth bound \cite{lyth,lyth2} on the minimum
variation in the inflaton field during 
inflation, namely, 
$\delta \phi> M_{Pl}\sqrt {r/7}\, \delta N$, where $\delta N$
is taken to be 
the number of e-foldings of inflation $( \sim 4.6)$ over which
modes corresponding to multipoles $l< 100$ leave the horizon.
This bound is further strengthened by a factor of 2.5 by
including the additional 7 e-foldings (for GUT scale inflation)
required by our analysis. The revision of the Lyth analysis by
including the entire duration of inflation \cite{lyth3} should
also include the additional e-foldings given in
Eq. (\ref{dN}).

 In conclusion, we find that if inflation is preceded by a
hot radiation era then there will be a stimulated emission of
inflaton perturbations during inflation into the  initial thermal
bath of the inflaton which will change the scale free spectrum.
The change in the spectrum is large at 
large angles and there is no change in the CMB anisotropies at
small angles.
(This differs from the warm inflation scenario where the departure from
scale invariance is proportional to the derivative of the inflaton dissipative
term, which is small to meet the slow roll condition \cite{warm2}.)
Furthermore, from the WMAP data we find that the minimum
number of e-foldings of inflation is larger than that required to solve the
horizon problem, with the increase varying from 7-32 for the energy scale of
inflation varying from the GUT to the electroweak scale.

\noindent {\bf Acknowledgement:} We thank M. S. Santhanam for helping
us with his computational skills.

\end{multicols}
\end{document}